\documentclass[aps,prb,twocolumn,showpacs,showkeys]{revtex4}
\usepackage{graphicx,psfrag}
\usepackage{mathptmx}
\usepackage{epstopdf}
\usepackage{hyperref}

\begin{document}


\title{Multiband effective bond-orbital model for nitride semiconductors with wurtzite structure}

\author{Daniel Mourad}
\email{dmourad@itp.uni-bremen.de}
\author{Stefan Barthel}
\author{Gerd Czycholl}
\affiliation{Institute for Theoretical Physics, University of
Bremen, D-28359 Bremen, Germany}



\date{\today}
\begin{abstract}
A multiband empirical tight-binding model for group-III-nitride semiconductors with a wurtzite structure has been developed and applied to both bulk systems and embedded quantum dots. As a minimal basis set we assume one $s$-orbital and three $p$-orbitals, localized in the unit cell of the hexagonal Bravais lattice, from which one conduction band and three valence bands are formed. Non-vanishing matrix elements up to second nearest neighbors are taken into account. These matrix elements are determined so that the resulting tight-binding band structure reproduces the known $\Gamma$-point parameters, which are also used in recent $\mathbf{k\cdot p}$-treatments. Furthermore, the tight-binding band structure can also be fitted to the band energies at other special symmetry points of the Brillouin zone boundary, known from experiment or from first-principle calculations. In this paper, we describe details of the parametrization and present the resulting tight-binding band structures of bulk GaN, AlN, and InN with a wurtzite structure. As a first application to nanostructures, we present results for the single-particle electronic properties of lens-shaped InN quantum dots embedded in a GaN matrix.
\end{abstract}


\maketitle


\section{Introduction\label{sec:intro}}
Due to their unique physical properties, zero-dimensional semiconductor nanostructures, realized by either epitaxial growth or colloidal chemical synthesis, \cite{guzelian_colloidal_1996} offer a broad range of applications.\cite{michler_single_2009} The three-dimensional confinement of spatially localized charge carriers in such tailor-made systems leads to a discrete and tunable one-particle spectrum, which can be used for a variety of optoelectronic applications, for quantum computing and quantum cryptography, and even for nanobiological applications like biological flourescence labeling.~\cite{bruchez_semiconductor_1998, michalet_quantum_2005}

Most binary group II-VI and III-V semiconductor materials and their ternary and quaternary alloys crystallize in the cubic zincblende or in the hexagonal wurtzite phase, so the corresponding nanostructures  can also be attributed to one of these two structures. Dependent on the material system and the adequate parameter range of the experimental conditions (e.g. growth temperature and substrate type for epitaxial growth, additionally the chemical environment and particle size for colloidal synthesis), it is even possible nowadays to realize either of the two structures for the same compounds. For example, epitaxially grown GaN/AlN quantum dots can be produced in the metastable zincblende modification  and in the thermodynamically stable wurtzite configuration.~\cite{lazar_investigation_2004}

The calculation of the optical properties of such systems requires the knowledge of a set of single particle eigenstates and eigenvalues for the confined carriers (electrons and holes), which can be obtained by means of different methods. Rather simple models like effective mass approximations \cite{grundmann_inas/gaas_1995, wojs_electronic_1996, shi_effects_2003} can give a first insight into the behaviour of such systems. Multiband $\mathbf{k}\cdot\mathbf{p}$-models \cite{fonoberov_excitonic_2003, pryor_eight-band_1998, stier_electronic_1999, andreev_theory_2000} incorporate higher effects like valence band mixing, but still make use of the envelope function approximation, thus not resolving the characteristic underlying lattice structure. Nevertheless, they have been succesfully applied  to various material systems and were extended by the inclusion of strain and piezoelectricity effects.

Empirical pseudopotential models (EPM) \cite{wang_pseudopotential_1996, wang_linear_1999, wang_comparison_2000, bester_cylindrically_2005} and empirical tight-binding models (ETBM) \cite{santoprete_tight-binding_2003, schulz_tight-binding_2005, schulz_electronic_2006, schulz_tight-binding_2006, schulz_spin-orbit_2008, korkusinski_building_2008}  allow for the possibility of a microscopic description of nanostructures. While the EPM is capable of resolving variations on the atomic scale, it requires a large set of basis states, which limits the application of these models to small nanostructures. The ETBM uses a coarse graining on the scale of lattice sites, which makes it possible to stick to a small set of basis states and perform calculations on larger supercells with feasible effort. Additionally, it gives a rather intuitive real-space picture of the system in terms of localized Wannier states. The coupling between different lattice sites is usually limited to first or second nearest neighbors, depending on the purpose. The goal is to analytically deduce a manageable set of equations for the tight-binding (TB) matrix elements in terms of bulk parameters  (e.g. the band gap, effective masses, spin-orbit splitting) which can be accessed  either from experiment or from first-principle calculations like DFT - LDA or recent $G_0W_0$-results.~\cite{rinke_band_2006, rinke_consistent_2008} These TB  matrix elements then enter the nanostructure calculation.


When establishing an empirical tight-binding model, one can start from a L\"owdin-orthogonalized atomic basis and use a linear combination of atomic orbitals (LCAO) as ansatz for the required eigenstates.~\cite{slater_simplified_1954} For nitride semiconductors with a wurtzite structure this has been done recently by Schulz \textit{et al}.~\cite{schulz_tight-binding_2006} But it is as well justified to start from ``effective bond-orbitals'', i.e. Wannier-like orbitals localized within a unit cell. Within the LCAO spirit, these effective orbitals can in principle be expressed as linear combinations of the above mentioned atomic orbitals. Neither the atomic orbitals nor the effective orbitals are explicitly known or required within an empirical TB approach, as only the matrix elements between those orbitals are needed to obtain the TB band structure. Therefore, it is equally justified to perform the parametrization directly for the effective bond-orbitals so that known bulk band structure properties are reproduced. Such a version of an ETBM is commonly called ``effective bond-orbital model`` (EBOM). The EBOM has the advantage that it usually allows for a better fit throughout the whole Brillouin zone (BZ) within a given basis set.

The EBOM has long been established for the cubic zincblende structure; a first EBOM parametrization by Chang \cite{chang_bond-orbital_1988} incorporated three-center overlap integrals in a basis set of one $s$- and three $p$-orbitals on each site of the fcc Bravais lattice. This parametrization
was restricted to coupling up to nearest neighbors, so that only the band energies at the  $\Gamma$-point were fitted, besides the usual set of effective conduction band masses and corresponding valence band parameters. Loehr augmented this model in Ref.  \onlinecite{loehr_improved_1994} by the inclusion of hopping up to second-nearest
neighbors to additionaly fit the band-structure of the bulk material to an extended parameter
set, including the $X$-point energies. This resulted in a better agreement of the resulting tight-binding conduction band with first-principle calculations.~\cite{fritsch_band_2004}

To our knowledge, there exists only one parametrization of the EBOM for materials with wurtzite structure in the literature.~\cite{chen_bond_2004} As this work is restricted to a nearest neighbor parametrization and a fit to zone center energies only, we developed a new parametrization including second nearest neighbor matrix elements and a fit to band energies at other special BZ points. We apply this EBOM to the calculation of the electronic properties of lens-shaped InN quantum dots embedded within GaN. 

This work is organized as follows. In Sec. II, our specific EBOM is presented. We developed  a second nearest neighbor parametrization. Furthermore, we describe the application to zero-dimensional nanostructures and discuss the inclusion of strain and piezoelectric fields. In Sec. III, the  one-particle spectrum for  a lens-shaped InN quantum dot embedded within GaN and a comparison with results from other $\mathbf{k}\cdot\mathbf{p}$- and fully microscopic ETBM calculations is given. Section IV contains a summary, a conclusion and a brief outlook to possible extensions of our model.

\section{Theory \label{sec:theory}}

\subsection{Effective bond-orbital model for bulk semiconductors \label{subsec:EBOMbulk}}

Linear combinations of atomic orbitals within one unit cell can be used as ansatz for the Wannier functions localized within the unit cell of the Bravais lattice. From these Wannier functions, the extended Bloch functions can be determined by means of a unitary transformation. As neither the atomic states nor the Wannier functions are explicitly used in an empirical tight-binding model, it is not necessary to start from the atomic wave functions, but one can directly assume a basis of Wannier-like functions, which is the basic idea of the EBOM approach.

As the conduction band wave functions at the BZ center predominantly transform $s$-like with some $p_z$ character, while the corresponding valence band wave functions transform like $p$-states with some $s$ character, we use a localized $sp^3$ basis per spin direction:
\begin{equation}\label{eq:sp3basis}
\left| \mathbf{R}, \alpha \right\rangle, \quad \alpha \in \left\lbrace s\uparrow,p_x\uparrow,p_y\uparrow,p_z\uparrow,s\downarrow,p_x\downarrow,p_y\downarrow,p_z\downarrow \right\rbrace .
\end{equation}
Here $\mathbf{R}$ labels the $N$ sites of the hexagonal lattice, which is the underlying Bravais lattice of the wurtzite crystal structure. 

A trial wave function that satisfies the Bloch condition is the Bloch sum 
\begin{equation}\label{eq:Blochsum}
\left| \psi_{\mathbf{k}} \right\rangle =   \frac{1}{\sqrt{N}} \sum_{\alpha} c_{\alpha }(\mathbf{k}) \sum_{\mathbf{R}} e^{i \mathbf{k} \cdot \mathbf{R}} \left| \mathbf{R}, \alpha \right\rangle.
\end{equation}
The band structure $E(\mathbf{k})$ is now given by the solution of the secular equation
\begin{equation}\label{eq:EBOMse}
\sum_{\alpha'} \, H_{\alpha \alpha'}(\mathbf{k}) \, c_{\alpha'}(\mathbf{k}) = E(\mathbf{k}) \, c_{\alpha}(\mathbf{k}),
\end{equation}
for each wave vector $\mathbf{k}$, where
\begin{equation} \label{eq:EBOMmatrix}
 H_{\alpha \alpha'}(\mathbf{k}) = \sum_{\mathbf{R},\mathbf{R'}} e^{i \mathbf{k} \cdot (\mathbf{R} - \mathbf{R'})} E_{\alpha  \alpha'}^{\mathbf{R} \mathbf{R'}}.
\end{equation}
The EBOM matrix elements  of the bulk Hamiltonian $H^{\text{bulk}}$ are thus given by
\begin{equation}\label{eq:EBOMme}
E_{\alpha \alpha'}^{\mathbf{R} \mathbf{R'}} = \left\langle \mathbf{R}, \alpha \right| H^{\text{bulk}} \left| \mathbf{R'}, \alpha' \right\rangle.
\end{equation}
It should explicitly be pointed out that the artificial change of point group symmetry from $C_{3v}$ (wurtzite) to $C_{6v}$ (hexagonal lattice) in the EBOM approach does \emph{not} uniquely stem from the omission of the atomic basis, but rather from the specific set of basis functions used. For instance, the original inversion asymmetry of the wurtzite crystal could be restored when the set of basis functions, Eq. (\ref{eq:sp3basis}), is extended by states that are not parity eigenstates. This has been done for cubic systems by Cartoix\`a \textit{et al.} in Ref. \onlinecite{cartoixa_description_2003}.

To include the influence of spin-orbit coupling, we follow Ref. \onlinecite{chadi_spin-orbit_1977}. As we expect
the spin-orbit part of  $H_{\text{bulk}}$ to be of weak influence, we assume only  site-diagonal contributions, which stem from the $p$-orbitals. Additionally, the non-ideal $c/a$ lattice constant ratio energetically seperates the $p_z$- from the $p_x$- and the $p_y$-orbitals. These effects can properly be incorporated by introduction of one spin-orbit splitting parameter $\Delta_{\text{so}}$ and one crystal field splitting parameter $\Delta_{\text{cr}}$.

When restricting the non-vanishing matrix elements, Eq. (\ref{eq:EBOMme}), up to nearest or second nearest neighbors, the secular equation (\ref{eq:EBOMse}) can be solved analytically for high symmetry points throughout the BZ of the hexagonal lattice. This yields a set of equations for the EBOM matrix elements in terms of the energetic positions of the bands at the critical $\mathbf{k}$-values. By expanding the elements of Eq. (\ref{eq:EBOMmatrix}) around the BZ center and comparing the matrix representation to a corresponding $\mathbf{k}\cdot\mathbf{p}$-Hamiltonian, ~\cite{winkelnkemper_interrelation_2006,chuang_kp_1996} it is possible to deduce additional constraints in terms of the conduction band effective masses and corresponding valence band parameters.

The goal is to arrive at a solvable set of equations which link a sufficiently large number of $E_{\alpha \alpha'}^{\mathbf{R} \mathbf{R'}}$ to a desired set of band structure parameters. In practice, this will require the additional omission of either matrix elements or band parameters, as the system of equations  becomes rather complicated.  The low symmetry of the hexagonal lattice will result in a larger number of independent parameters and equations than in the case of cubic crystal systems. An overview of the results for a coupling up to second nearest neighbors is given in Tab. \ref{tab:parametrizations}.
More details on the parametrization are given in App. \ref{appA}.
\begin{table}
	\caption{Overview of the EBOM parametrization with coupling up to second nearest neighbors. The nomenclature for the band energies follows the usual single group notation, see e.g. Ref. \onlinecite{yeo_electronic_1998}. Note that an additional valence band parameter $A_7$ has been neglected.}
	\label{tab:parametrizations}
	\begin{tabular}{ll}
		\hline
		\hline
		\textbf{Second nearest neighbor coupling:} & 26 EBOM matrix elements\\
		\hline
		\hline
		Band parameter & Description \\
		\hline
		$E_g = \Gamma_1^c - \Gamma_6^v$ & direct band gap at $\Gamma$ \\
		$m_e^\parallel$, $m_e^\perp$ & effective electron masses \\
		$A_1$, $A_2$, $A_3$, $A_4$, $A_5$, $A_6$ & valence band parameters  \\
		$\Delta_{so}$ & spin-orbit splitting \\
		$\Delta_{cr}$ & crystal field splitting \\
		$A_{1,3}^c$, $A_{5,6}^v$, $A_{1,3}^v$ & $A$-point energies \\
		$L_{1,3}^c$, $L_{1,3}^v$, $L_{2,4}^v$, $L_{1,3^\prime}^v$ & $L$-point energies\\
		$M_{1}^c$, $M_{4}^v$, $M_{3}^v$, $M_{1}^v$ & $M$-point energies\\
		$H_{3}^c$, $H_{3}^v$, $H_{3^\prime}^v$ & $H$-point energies\\		
		$E_p^{\parallel,\perp} = f(E_g, \Delta_{so},\Delta_{cr}, m^{\parallel,\perp})$ & Kane parameters \\
		\hline
		\hline
	\end{tabular}
\end{table}

The band structure for this parametrization  is now obtained by the diagonalization of the $8 \times 8$ matrix $H_{\alpha \alpha'}(\mathbf{k})$, Eq. (\ref{eq:EBOMmatrix}), for each $\mathbf{k}$.

For all calculations in the present paper, two distinct parameter sets have been used. The first parameter set is derived from a consistent set of band
parameters \cite{rinke_consistent_2008} obtained from $G_0W_0$ calculations based on exact-exchange
optimized effective potential ground states (OEPx),~\cite{rinke_combining_2005} supplemented by
additional band structure energies at high symmetry points.~\cite{rinke_personal_2009} Since the $G_0W_0$@OEPx band gaps and crystal field splittings still differ slightly from the
experimental values, we decided to use a second parameter set, in which these values were replaced by the parameters recommended by Vurgaftman and Meyer in 2003.~\cite{vurgaftman_band_2003} To obtain the correct band gaps, all conduction band energies from the $G_0W_0$ calculations were shifted by the respective difference in the second parameter set. Moreover, we used the spin-orbit splittings of Ref. \onlinecite{vurgaftman_band_2003} in both sets, as the $G_0W_0$ calculations did not include the electron spin. This should be a reasonable approach, as the spin-orbit splitting is comparatively small in these systems. The two parameter sets are listed in Tab. \ref{tab:parametersets} and will be referred to simply as ''$G_0W_0$ parameters`` and "corrected $G_0W_0$ parameters'' from now on.

In our opinion, the corrected $G_0W_0$ parameters should clearly be preferred, as it is known  that even highly sophisticated ab-initio approaches still do not properly reproduce the band gap.

\begin{table}
\caption{Empirical parameter sets used in the EBOM calculations. The first parameter set  corresponds to results obtained in a DFT + $G_0W_0$ treatment by Rinke \textit{et al.},~\cite{rinke_consistent_2008, rinke_personal_2009} while the second parameter set  replaces some parameters by values recommended by Vurgaftman \textit{et al.}.~\cite{vurgaftman_band_2003} See the text for further discussion. Blank cells mean the adoption of the parameter of the alternate set.}
\label{tab:parametersets}
	\begin{tabular}{c|ccc|ccc}
	\hline
	\hline
	Reference & \multicolumn{3}{|c}{$G_0W_0$ parameters} & \multicolumn{3}{|c}{Corrected $G_0W_0$ parameters}\\
	\hline
	Material: & AlN & GaN & InN & AlN & GaN & InN \\
	\hline
	$a$ [\AA] & 3.110 & 3.190 & 3.540 & & & \\
	$c$ [\AA] & 4.980 & 5.189 & 5.706 & & & \\
	\hline
	$E_g$ [eV] & \phantom{-}6.464 & \phantom{-}3.239 & \phantom{-}0.694 & \phantom{-}6.250 & \phantom{-}3.510 & \phantom{-}0.78 \\
	$\Delta_{so}$ [eV] & & & & \phantom{-}0.019 & \phantom{-}0.017 & \phantom{-}0.005 \\
	$\Delta_{cr}$ [eV] & -0.295 & \phantom{-}0.034 & \phantom{-}0.066 & -0.169 & \phantom{-}0.010 & \phantom{-}0.040 \\
	$E_p^{\parallel}$  [eV] & 16.972\footnotemark[1] \footnotetext[1]{For a given band gap, $E_p^{\parallel,\perp}$ are not independent parameters when $ m^{\parallel,\perp}$ are known (see Eqs. (\ref{eq:E_p}) - (\ref{eq:P_perp})) in App. \ref{appA}. To obtain a better fit to the $G_0W_0$- band structure, these parameters can be adjusted by least-square fit values, see Ref. \onlinecite{rinke_consistent_2008}. When  the band gap is subsequently altered, as in the set to the right, the analytic expression has to be used again.} & 17.292\footnotemark[1]  &  \phantom{-}8.742\footnotemark[1] & \multicolumn{3}{c}{$ f(E_g, \Delta_{so},\Delta_{cr}, m^{\parallel})$}  \\
	$E_p^{\perp}$ [eV] & 18.165\footnotemark[1] & 16.265\footnotemark[1] &  \phantom{-}8.809\footnotemark[1] & \multicolumn{3}{c}{$ f(E_g, \Delta_{so},\Delta_{cr}, m^{\perp})$} \\
	\hline
	$m_e^\parallel$ [$m_0$] & \phantom{-}0.322\footnotemark[1] & \ 0.186\footnotemark[1] & \ 0.065\footnotemark[1] & & & \\
	$m_e^\perp$ [$m_0$] & \phantom{-}0.329\footnotemark[1] & \ 0.209\footnotemark[1] & \ 0.068\footnotemark[1] & & & \\
	$A_1$  & -3.991& -5.947 & -15.803 & & & \\
	$A_2$  & -0.311 & -0.528 & -0.497 & & & \\
	$A_3$  & \phantom{-}3.671 & \phantom{-}5.414 & 15.251 & & & \\
	$A_4$  & -1.147 & -2.512 & -7.151 & & & \\
	$A_5$  & -1.329 & -2.510 & -7.060 & & & \\
	$A_6$  & -1.952 & -3.202 & -10.078 & & & \\
	\hline
	$A_{1,3}^c$ [eV] & \phantom{-}8.844 & \phantom{-}5.701 & \phantom{-}3.355 & \phantom{-}8.631 & \phantom{-}5.972 & \phantom{-}3.441 \\
	$A_{5,6}^v$ [eV] & -0.686 & -0.597 & -0.509 & & & \\
	$A_{1,3}^v$ [eV] & -3.573 & -4.110 & -3.581 & & & \\
	$L_{1,3}^c$ [eV] & \phantom{-}7.545 & \phantom{-}5.798 & \phantom{-}4.356 & \phantom{-}7.332 & \phantom{-}6.069 & \phantom{-}4.442 \\
	$L_{1,3}^v$ [eV] & -1.515 & -2.065 & -1.732 & & & \\
	$L_{2,4}^v$ [eV] & -1.689 & -2.144 & -1.838 & & & \\
	$L_{1,3^\prime}^v$ [eV] & -6.033 & -6.984 & -5.769 & & & \\
	$M_{1}^c$ [eV] & \phantom{-}8.084 & \phantom{-}6.550 & \phantom{-}4.934 & \phantom{-}7.870 & \phantom{-}6.821 & \phantom{-}5.020 \\
	$M_{4}^v$ [eV] & -0.837 & -1.111 & -0.997 & & & \\
	$M_{3}^v$ [eV] & -1.893 & -2.382 & -1.889 & & & \\
	$M_{1}^v$ [eV] & -3.649 & -4.518 & -3.714 & & & \\
	$H_{3}^c$ [eV] & \phantom{-}9.774 & \phantom{-}7.982 & \phantom{-}6.281 & \phantom{-}9.560 & \phantom{-}8.253 & \phantom{-}6.367 \\
	$H_{3}^v$ [eV] & -0.914 & -1.609 & -1.401 & & & \\
	$H_{3^\prime}^v$ [eV] & -5.202 & -6.474 & -5.422 & & & \\
	\hline
	\hline
	\end{tabular}
\end{table}

\begin{figure*}
	\includegraphics[width = \linewidth]{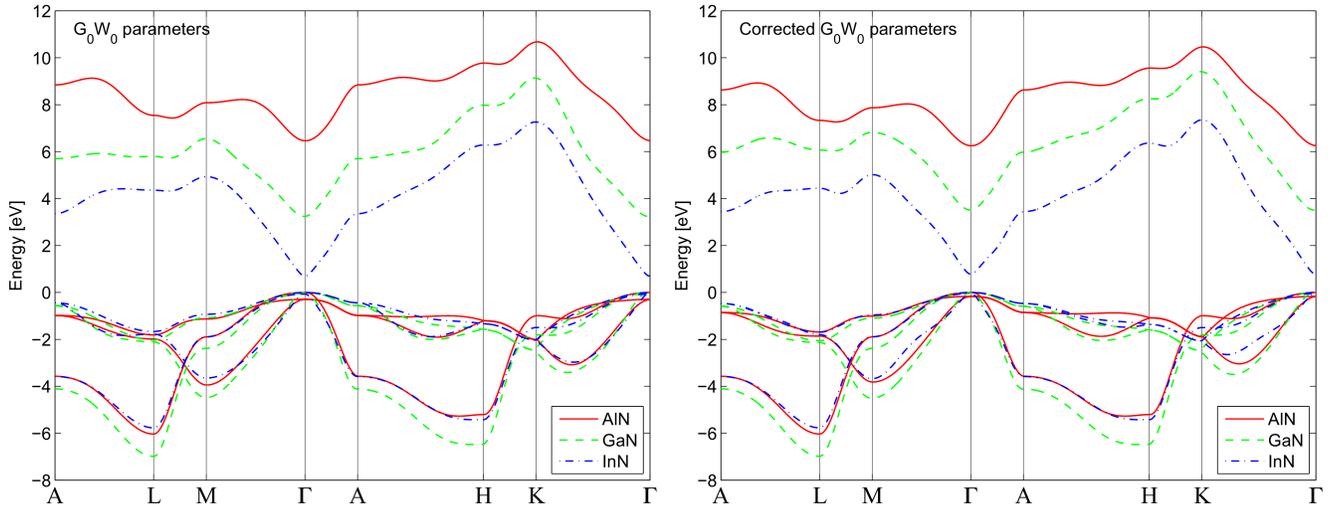}
\caption{(Color online) EBOM band structures for AlN, GaN and InN with coupling up to second nearest neighbors, using the $G_0W_0$ parameters (left image) and the corrected $G_0W_0$ parameters (right image). Further details are given in Tab. \ref{tab:parametersets} and in the text. The top of the valence band is set to zero, respectively.}
\label{fig:bandstructures}
\end{figure*}

The resulting band structures are depicted in Fig. \ref{fig:bandstructures} for AlN, InN and GaN. The top of the valence band of each material is set to zero. One can easily identify the direct band gap in the Brillouin zone center, one spin-degenerate conduction band and three spin-degenerate valence bands, according to the employed basis set, Eq. (\ref{eq:sp3basis}), of four orbitals per spin direction. The twofold Kramers degeneracy of each energy level $E(\mathbf{k})$  is a direct consequence of the time-reversal symmetry, as no external magnetic field is applied. Due to the fitting to the multiple high symmetry points on the BZ surface, each band has a finite bandwidth of a realistic magnitude, which the $\mathbf{k}\cdot\mathbf{p}$-theory, of course, does not reproduce, as it is restricted to the vicinity of the BZ center within this basis set. In addition, no erroneous curvature of the bands into the band gap occurs for larger $|\mathbf{k}|$. 

\begin{figure}
	\includegraphics[width = \linewidth]{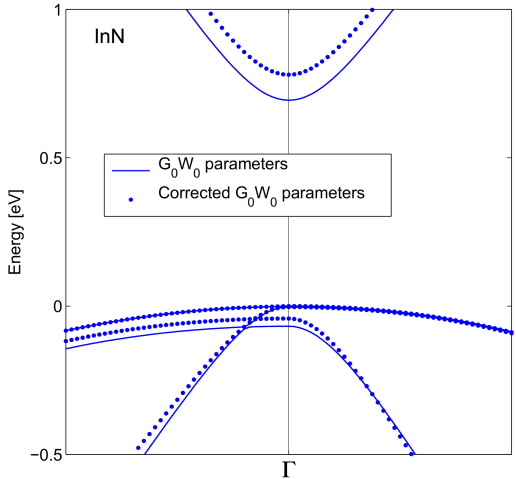}
\caption{(Color online) EBOM band structures for InN around $\Gamma$ for the two parameter sets. Further details are again given in Tab. \ref{tab:parametersets} and in caption of Fig. \ref{fig:bandstructures}.}
\label{fig:bandinlet_InN}
\end{figure}

At first glance, the band structures do not differ significantly for both parameter sets. To emphasize the differences, Fig. \ref{fig:bandinlet_InN} shows the band structure of InN around the $\Gamma$-point. By having a closer look, one can see that the energetic positions are different, because of the different crystal field splitting and band gap. Also, the curvatures of the bands differ for the two parameter sets. This is a result of the slightly different $E_p^{\parallel,\perp}$ and will be adressed again in Sec. \ref{subsec:spspInN}.   More sophisticated methods for band structure calculation will give more conduction and valence bands in the energetic range around the band gap. Although this feature could also be included in our EBOM approach by augmenting the number of orbitals per unit cell, it would not only result in a more complicated parametrization, but also lead to significantly higher computational costs for nanostructure calculations. Thus we stick to a minimal basis set of four bands per spin direction, which gives a reasonable agreement with the ''true`` band structure in the region of interest. The reliability of this basis set has also been established by the variety of existing eight-band-$\mathbf{k}\cdot\mathbf{p}$ calculations for these systems and comparisons to experimental results for device applications.~\cite{winkelnkemper_interrelation_2006, winkelnkemper_polarized_2007}

Of course, our EBOM parametrization is not limited to a specific set of parameters. The fit to the energies at the BZ boundaries allows for an adaption to a wide range of parameters, as these additional constraints practically prevent spurious solutions where the bands curve into the band gap far away from the BZ center. These problems are widely known to occur for simpler $\mathbf{k}\cdot\mathbf{p}$- and tight-binding parametrizations when an inappropriate set of $\Gamma$-point parameters is used.

With our model, a more systematic investigation of the influence of single parameters on the properties of low-dimensional systems is possible, as the band structure is not sensitive to small perturbations in the input parameters. This stability of the parametrization transfers directly to the application on nanostructures, as e.g. spurios solutions in the bulk band gap will lead to corresponding states in the forbidden energy region of the nanostructure.
In spite of the progress in the field of both sophisticated ab-initio calculations and highly refined experiments, this is and will remain an important feature, as certain physical quantities like band offsets, Luttinger parameters and optical matrix elements remain  ambiguous because they are only indirectly measurable and depend on  model assumptions.

%
%

\subsection{Application of the EBOM to quantum dots \label{subsec:EBOMQDs}}
As we now have  determined the EBOM matrix elements for the bulk materials, they can be used as input in the calculations for a quantum-confined nanostructure. In case of a quantum dot, the translational invariance is lost in all three spatial dimensions, so the adequate ansatz for an eigenstate, Eq. (\ref{eq:Blochsum}), is reduced to a direct linear combination of localized effective orbitals
\begin{equation}
\left| \psi \right\rangle =  \sum_{\alpha,\mathbf{R}} c_{\mathbf{R} \alpha } \left| \mathbf{R}, \alpha \right\rangle.
\end{equation}
The corresponding secular equation is now given by
\begin{equation}\label{eq:seceqQD}
\sum_{\alpha', \mathbf{R'} } E_{\alpha \alpha' }^{\mathbf{R}\mathbf{R'}} \, c_{\mathbf{R'} \alpha'} = E \, c_{\mathbf{R} \alpha},
\end{equation}
where $E_{\alpha \alpha' }^{\mathbf{R}\mathbf{R'}}$ are the EBOM matrix elements from Eq. (\ref{eq:EBOMme})  and the site indices $\mathbf{R}, \mathbf{R'}$ now range over the finite $N$ sites of a sufficiently large supercell. In our present $sp^3$-basis, the eigenstates and eigenenergies of Eq. (\ref{eq:seceqQD}) are obtained as the solutions of a $8N \times 8N$ matrix eigenvalue problem.
According to Refs. \onlinecite{schulz_tight-binding_2006} and \onlinecite{schulz_tight-binding_2005}, a nanostructure made of one material A embedded in a barrier of material B can be modelled by using the matrix elements of the A-material for the corresponding lattice sites and vice versa. For the interface, a linear interpolation of the corresponding hopping matrix elements is used. The confinement potential for the carriers can properly be incorporated by an upward shifting of the diagonal elements of the A-material by the valence band offset $\Delta E_v$ between the two materials. When the band gap between the materials B and A exceeds this offset, we are naturally left with a type-I confinement potential for the electrons and holes.

The  application to one- or two-dimensional structures is a trivial task and can be done correspondingly. 

\subsection{Possible inclusion of piezoelectricity and strain\label{subsec:piezoandstrain}}

In the systems under consideration, there is always a spontaneous polarization due to the deviation of the $c/a$-ratio from the value in the ideal wurtzite structure.
Additionally, strain fields will influence the electronic properties of these systems, if present.
While there are in fact zero-dimensional systems, like fully relaxed nanocrystals, where this effect can be neglected, the below presented model system of epitaxially grown InN quantum dots embedded in a GaN matrix will in fact be strained due to the lattice mismatch, and this strain will not only shift the band edges, but also alter the equilibrium positions of the lattice sites and thus the piezoelectric charge density.

Both, the spontaneous and the strain induced polarization can be incorporated into the tight-binding calculations by the solution of the Poisson equation. \cite{schulz_tight-binding_2006,schulz_electronic_2006}
 Although it has been discussed in previous publications like Ref. \onlinecite{schulz_tight-binding_2006}, that for this specific quantum dot system a proper inclusion of a constant band-edge shift might be sufficient, a more general approach is of course desirable.

Again, the one-to-one correspondence to the $\mathbf{k}\cdot\mathbf{p}$ model at $\Gamma$ allows for a straight-forward inclusion of strain effects on the bulk band structure by augmenting the $\mathbf{k}\cdot\mathbf{p}$-Hamiltonian by a strain-dependant part, as done in Refs. \onlinecite{winkelnkemper_interrelation_2006} and \onlinecite{chuang_kp_1996}.
The then obtained analytical dependance of the EBOM matrix elements $E_{\alpha \alpha'}^{\mathbf{R} \mathbf{R'}}$ on the deformation potentials $a_{1,2}$ of the conduction band, $D_{1-6}$ of the valence bands and the elastic stiffness constants can then be used either to determine a distance-dependant scaling law for the $E_{\alpha \alpha'}^{\mathbf{R} \mathbf{R'}}$(similar to the famous Harrison $d^{-2}$ ansatz \cite{froyen_elementary_1979}) or directly be incorporated into the nanostructure Hamiltonian. In both cases, additionally an appropriate strain field has to be calculated for the low-dimensional system under consideration, either by atomistic \cite{fonoberov_excitonic_2003, andreev_theory_2000} or continuum mechanical \cite{winkelnkemper_interrelation_2006} approaches.

As this extension of the EBOM requires a careful comparison to further experimental and theoretical results, it is a topic of its own and part of ongoing research. Therefore, it will not  furtherly be adressed in the present publication. In the following section, we will neglect the influence of piezoelectricity and strain, in order to focus on the direct influence of the slightly different Kane parameters $E_p^{\parallel,\perp}$ on the single particle results. 

\section{Results for quantum dots \label{sec:results}}
\subsection{Model quantum dot geometry \label{subsec:modelQD}}

\begin{figure}
	\includegraphics[width = \linewidth]{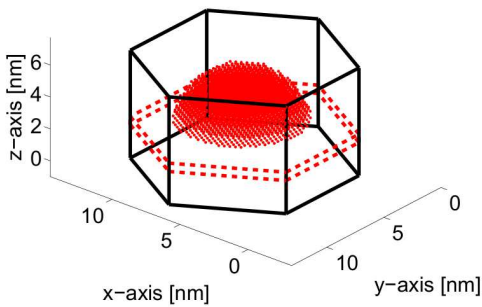}
\caption{(Color online) Geometry for the lens-shaped InN quantum dot on an InN wetting layer, embedded in GaN. The InN lattice sites of the QD are depicted with red dots, additionally, the intersections of the wetting-layer with the boundaries of the hexagonal supercell are visualized by the dotted red lines.}
\label{fig:geometry}
\end{figure}

Earlier ETBM calculations of Refs. \onlinecite{schulz_tight-binding_2006}, \onlinecite{schulz_electronic_2006} for the InN/GaN material system were performed using cubic supercells and fixed boundary conditions. In this paper we present a different and improved kind of supercell, which is depicted in Fig. $\ref{fig:geometry}$, by means of keeping the point group symmetry of the underlying hexagonal Bravais lattice in combination with periodic boundary conditions. This leads to several benefits, e.g. no artificial surface states can arise in the single-particle spectrum, and in contrast to cubic supercells our hexagonal one does not interfere with the $C_{6v}$ point group symmetry of the lattice. The simulated lens-shaped InN quantum dot has a diameter of 7.7 nm and a height of 3.1 nm. It is placed on top of a wetting-layer with thickness of one $c$ lattice constant, since Stranski-Krastanov growth-mode is assumed for the given structure. The surrounding GaN supercell has a dimension of $ 36.4\, a \times 42\, a \times 14\, c$ with respect to the cartesian axes. With this size, convergence for the one-particle wave functions is ensured. Furthermore, a completely strained structure is supposed, so that no deviations from the ideal lattice positions emerge. As valence band offset, we use the value recommended by Vurgaftman \textit{et al.} \cite{vurgaftman_band_2003} of $\Delta E_v = 0.5$ eV for both parameter sets.

\subsection{One-particle spectrum for embedded InN quantum dot\label{subsec:spspInN}}
The numerical diagonalization of the corresponding nanostructure Hamiltonian (using the \textit{folded spectrum method}\cite{wang_solving_1994})  gives the desired  single-particle states and eigenenergies around the energy gap of the quantum dot. We solve Eq. (\ref{eq:seceqQD}) for eight bound electron and hole states, using the EBOM parametrization for second nearest neighbors and  taking spin-orbit coupling and crystal field splitting into account.

\begin{figure*}
	$G_0W_0$ parameters
	\includegraphics[width = \linewidth]{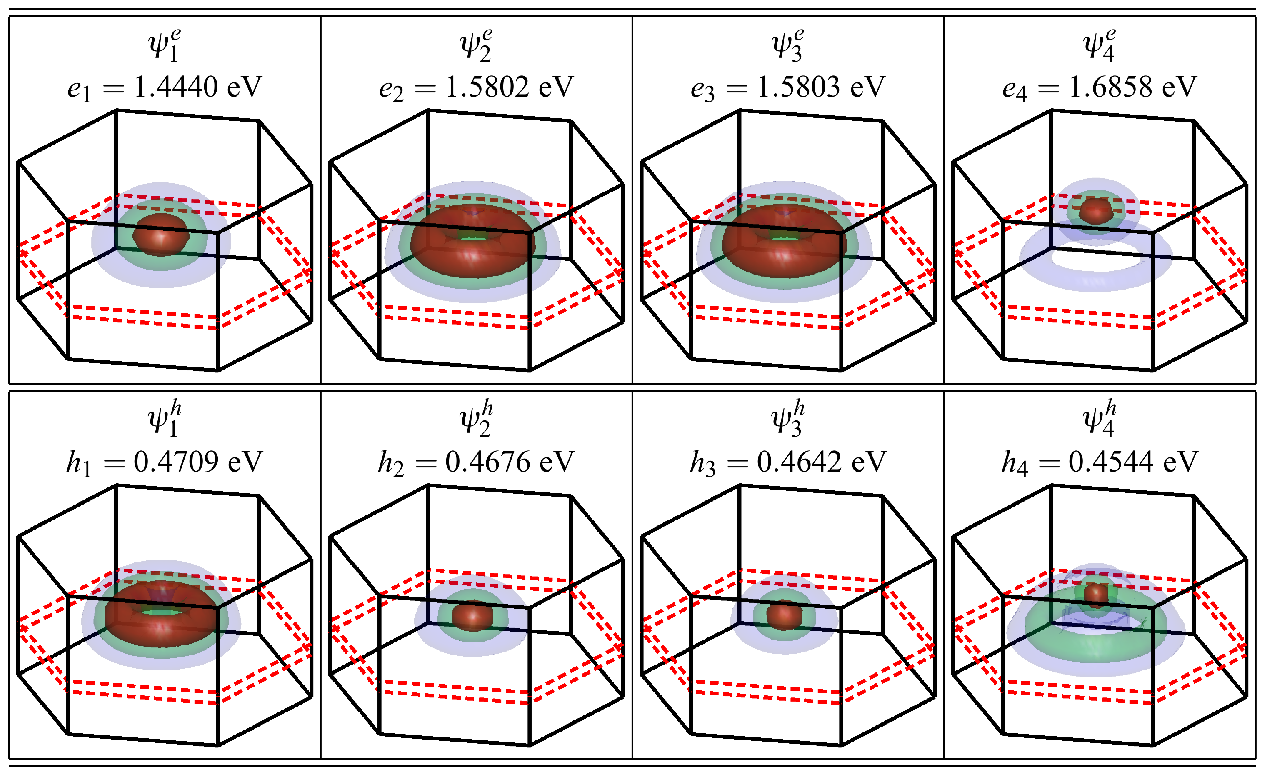}
	Corrected $G_0W_0$ parameters
	\includegraphics[width = \linewidth]{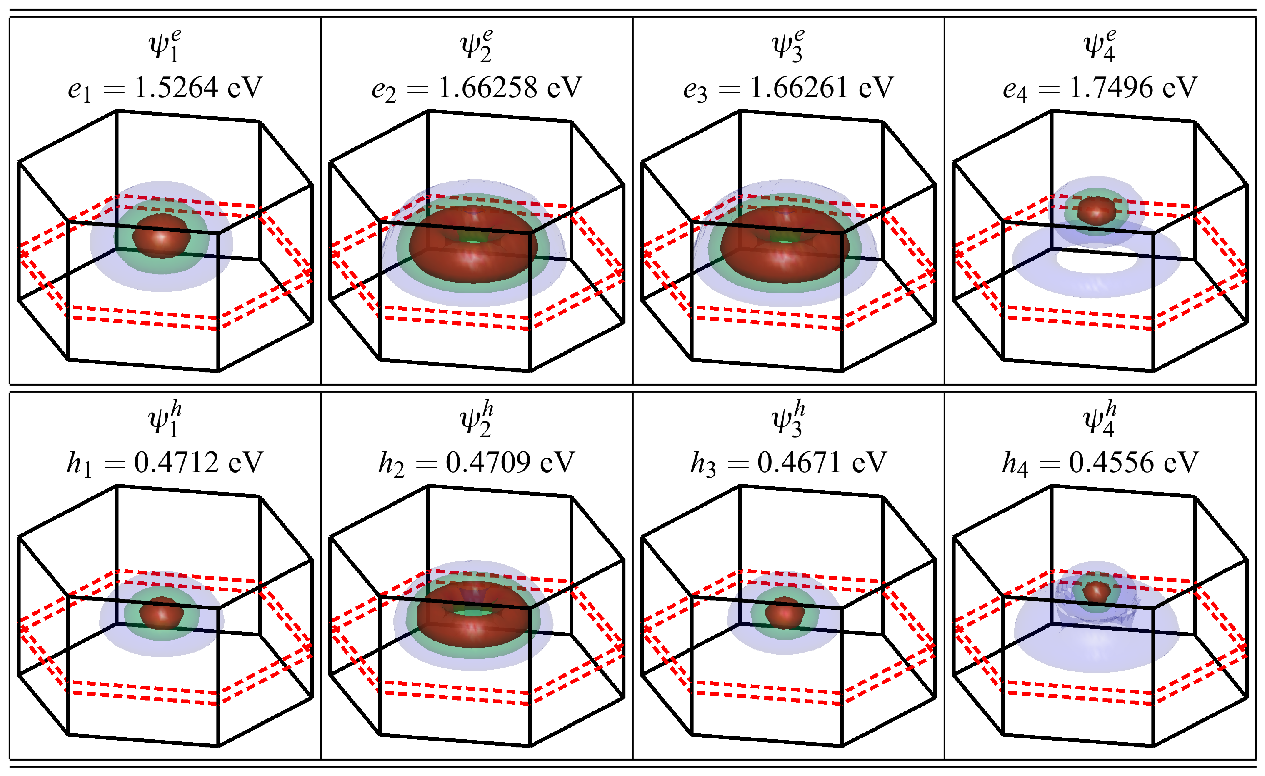}
\caption{(Color online) Visualization of the probability density by isosurfaces of 75\% (red), 45\% (green) and 15\% (blue) of the maximum value for electron $\psi_i^e$ and hole $\psi_i^h$ states within the hexagonal supercell. The red dotted lines give the intersection of the wetting layer with the cell boundary.
 In addition, the corresponding one-particle eigenenergies are shown. All energies are given with respect to the valence band edge of GaN. The two parameter sets have been presented in Tab. \ref{tab:parametersets}.  Please note the interchanging of the states $h_1$ and $h_2$ when using the latter set of parameters.}
\label{fig:propdens}
\end{figure*}

The resulting eigenfunctions are visualized in Fig. \ref{fig:propdens} by isosurfaces of the probability density, supplemented by the respective eigenenergies. All states are invariant under rotations by $\frac{\pi}{3}$ around the growth direction, according to the $C_{6v}$ point group symmetry of the Bravais lattice.
Each state is once again twofold degenerate due to time-reversal symmetry and well localized within the InN quantum dot. Tab. \ref{tab:orbitalcoefficients} reveals that the electron states mainly stem from the $s$-like conduction band, so a classification by their nodal structure is possible. $\psi_1^e$ is $s$-like, while $\psi_2^e$ and $\psi_3^e$ are complex linear combinations of the form $p_{\pm}=\frac{1}{\sqrt{2}}(p_x\pm ip_y)$. $\psi_4^e$ is a $p_z$-like state, but distorted by the shape of the quantum dot. The hole states show similar transformation properties at first glance. Nevertheless, Tab. \ref{tab:orbitalcoefficients} reveals that at least two atomic $p$-states contribute to their formation, so that they underlie strong band-mixing effects. This is in agreement with results from several other multiband calculations.~\cite{wei_valence_1996, fonoberov_excitonic_2003}

\begin{table}
		\begin{tabular}{c|c|c|c|c|c|c|c|c}
		\multicolumn{9}{c}{$G_0W_0$ parameters}\\
			\hline
			\hline
			& $e_1$ & $e_2$ & $e_3$ & $e_4$ & $h_1$ & $h_2$ & $h_3$ & $h_4$\\
			\hline
			$ s$ & \textbf{0.878} & \textbf{0.839} & \textbf{0.839} & \textbf{0.843} & 0.000 & 0.004 & 0.002 & 0.001\\
			$ p_x $ & 0.033 & 0.056 & 0.056 & 0.034 & \textbf{0.499} & \textbf{0.492} &  \textbf{0.494} & \textbf{0.495}\\
			$ p_y $ & 0.033 & 0.056 & 0.056 & 0.034 & \textbf{0.499} & \textbf{0.492} &  \textbf{0.494} & \textbf{0.495}\\
			$ p_z $ & 0.057 & 0.049 & 0.049 & 0.088 & 0.001 & 0.012 & 0.010 & 0.009\\
			\hline
			\hline
		\end{tabular}
		\begin{tabular}{c|c|c|c|c|c|c|c|c}
		\multicolumn{9}{c}{Corrected $G_0W_0$ parameters}\\
			\hline
			\hline
			& $e_1$ & $e_2$ & $e_3$ & $e_4$ & $h_1$ & $h_2$ & $h_3$ & $h_4$\\
			\hline
			$ s $ & \textbf{0.858} & \textbf{0.812} & \textbf{0.812} & \textbf{0.818} & 0.005 & 0.000 & 0.004 & 0.001\\
			$ p_x $ & 0.036 & 0.063 & 0.063 & 0.032 & \textbf{0.490} & \textbf{0.497} &  \textbf{0.491} & \textbf{0.490}\\
			$ p_y $ & 0.036 & 0.063 & 0.063 & 0.032 & \textbf{0.490} & \textbf{0.497} &  \textbf{0.491} & \textbf{0.490}\\
			$ p_z $ & 0.070 & 0.063 & 0.063 & 0.118 & 0.014 & 0.003 & 0.015 & 0.019\\
			\hline
			\hline
		\end{tabular}
\caption{Orbital contributions of the $sp^3$ basis to the bound one-particle states, calculated by  summation over the site and the spin index of the projections $\langle  \mathbf{R}, \alpha \vert \psi \rangle$. Dominant parts are marked in bold.}
\label{tab:orbitalcoefficients}
\end{table}

By taking a closer look at the degeneracies in Fig. \ref{fig:propdens}, we notice  no fourfold degeneracy of $\psi_2^e$ and $\psi_3^e$, in contrast to  $\mathbf{k}\cdot\mathbf{p}$-calculations from Ref. \onlinecite{winkelnkemper_interrelation_2006}, but reproduce the findings of earlier ETBM calculations from Ref. \onlinecite{schulz_spin-orbit_2008}, showing a twofold degeneracy for each bound state. The latter reference also includes a detailed group theoretical dicussion of this issue. The only deviations we can report on is the fact that the EBOM eigenenergies  of the electrons are more strongly bound for both parameter sets, compared to other ETBM results from Refs. \onlinecite{schulz_tight-binding_2006} and \onlinecite{schulz_electronic_2006}. This discrepancy can safely be attributed to the once again different set of input parameters and thus does not contradict comparative studies from Refs. \onlinecite{marquardt_comparison_2008} and \onlinecite{schulz_multiband_2009} for zincblende nanostructures, which show a very good agreement between these approaches. Our results for the hole  energies agree well with the earlier calculations of Refs. \onlinecite{schulz_tight-binding_2006} and \onlinecite{schulz_electronic_2006}  if  the piezoelectric field
is neglected (see e.g. Ref. \onlinecite{schulz_tight-binding_2006} for details).

A striking feature is the different order of the hole levels when switching between the parameter sets. The torus-shaped probability density belongs to the hole ground state $\psi^h_1$ with the $G_0W_0$ parameters, but is found as the first excited hole state $\psi^h_2$ when using the corrected $G_0W_0$ parameters. Another look at  Tab. \ref{tab:parametersets} reveals that the band gap of the dot material  differs by less than $90$ meV between these parameter sets; the confinement potential for the holes is even identical in both cases, as the same valence band offset is used. The crystal field splitting only differs by ca. $20$ meV for both materials. Obviously, this rather small variation of the bulk band gap and the crystal field splitting, which also results in slightly different Kane parameters $E_p^{\parallel,\perp}$ (see App. \ref{appA}), suffices to change the level structure. Further studies (not shown) reveal that even variations in the order of magnitude of the accuracy of the input parameters can change the order. In addition, the differing bulk band gaps of the two parameter sets lead to slightly  different one-particle energy gaps $E_g^{QD} = e_1-h_1$ of $\sim 0.97$ eV ($G_0W_0$ parameters) and $\sim 1.05$ eV (corrected $G_0W_0$ parameters), respectively.
When calculating optical properties like the excitonic absorption spectrum from the tight-binding single-particle spectrum, as e.g. done in Refs. \onlinecite{schulz_tight-binding_2006} and \onlinecite{schulz_multiband_2009}, the level structure and $E_g^{QD}$ are important characteristics. A change in the first will give rise to a change of the respective dipole matrix elements between the electron and hole states and thus alter the line intensity, while a change in $E_g^{QD}$ directly shifts the energetic position of the line itself. 

As the variation of the dot size and the proper implementation of strain effects and electrostatic built-in fields can additionally alter the level ordering, the influence of different material parameters should be carefully investigated when discussing the optical selection rules for a given geometry. For the system under consideration, we recommand the use of the corrected $G_0W_0$ parameters.

%

%
%

\section{Conclusion and outlook \label{sec:concl}}

In this paper, we have presented a multiband empirical tight-binding parametrization for both bulk semiconductors and nanostructures with a wurtzite structure, which represents an adaption of the effective bond-orbital model (EBOM) for the hexagonal phase.  A basis set of one $s$- and three $p$-orbitals for each spin direction is placed on the sites of the underlying hexagonal Bravais lattice.  Coupling up to second nearest neighbors has been used to fit one conduction and three valence bands to the energies and curvatures at the $\Gamma$-point and, additionally, to the energies at high symmetry points throughout the whole Brillouin zone.  The resulting band structures of the III-V-compounds InN, GaN and AlN were shown for two  disctinct parameter sets, namely the newest $G_0W_0$- results by Rinke \textit{et al.} and a slightly modified set, in which we adjusted single critical parameters by replacing them by the values given by Vurgaftman \textit{et al.} in order to obtain better agreement to experimental results.

In addition, we demonstrated the application of this parametrization to low-dimensional structures. A lens-shaped InN quantum dot on an InN wetting layer, embedded in a GaN matrix, has been modelled within a hexagonally shaped supercell with periodic boundary conditions. We have compared the resulting one-particle spectrum and the corresponding eigenstates to previous tight-binding results and have found a good concordance within the framework of the respective model. Furthermore, we have found that the two parameter sets yield a different order of hole states for the given dot diameter, although the corresponding bulk band structures barely differ at first glance. We strongly approve a careful review of the set of material parameters used in such calculations. For the present  InN/GaN quantum dot system, we recommend the use of the corrected $G_0W_0$ parameter set.

Besides the application to other quantum dot systems, like GaN in AlN, or different geometries, like coupled QDs or spherical nanocrystals,  the present parametrization can easily be applied to one-dimensional (quantum wires) or two-dimensional (quantum wells and superlattices) structures. Moreover, the effects of strain and piezoelectric built-in fields can be incorporated on different levels of sophistication, as suggested in the second  section of this publication.\\

\begin{acknowledgments}
The authors would like to thank Patrick Rinke for the provision of additional $G_0W_0$ results at further high symmetry points of the Brillouin zone.  Furthermore, we thank Stefan
Schulz for fruitful discussions. This work has been supported
by the Deutsche Forschungsgemeinschaft (research group
“Physics of nitride-based, nanostructured, light-emitting
devices”, project Cz 31/14-3).
\end{acknowledgments}


\appendix
\section{EBOM parametrization for the hexagonal lattice \label{appA}}

The  analytical dependance of the parameters
\begin{equation}
P_{\parallel,\perp} = \sqrt{\frac{\hbar^2}{2 m_0} E_{p}^{\parallel,\perp}}
\label{eq:E_p}
\end{equation}
of the band gap $E_g$, the spin-orbit and crystal field splittings $\Delta_{so}, \Delta_{cr}$ and the effective masses $m_e^{\parallel,\perp} $  is given by the following two equations:~\cite{chuang_kp_1996}
\begin{equation}
P_\parallel^2 =\frac{\hbar^2}{2m_0}\left(\frac{m_0}{m_e^\parallel}-1\right)\frac{3E_g(\Delta_{so}+E_g)+\Delta_{cr}(2\Delta_{so}+3E_g)}{2\Delta_{so}+3E_g},
\label{eq:P_par}
\end{equation}
\begin{equation}
P_\perp^2=\frac{\hbar^2}{2m_0}\left(\frac{m_0}{m_e^\perp}-1\right)E_g\frac{[3E_g(\Delta_{so}+E_g)+\Delta_{cr}(2\Delta_{so}+3E_g)]}{\Delta_{cr}\Delta_{so}+3\Delta_{cr}E_g+2\Delta_{so}E_g+3E_g^2}.
\label{eq:P_perp}
\end{equation}
The EBOM parametrization scheme with coupling up to second nearest neighbors gives a set of  equations which link the parameters of Tab. \ref{tab:parametersets} to the EBOM matrix elements $E_{\alpha \alpha'}^{\mathbf{R} \mathbf{R'}}$ of Eq. \ref{eq:EBOMme}. 
To obtain the desired number of free parameters for a one-to-one correspondance, one has to apply an adequate decomposition of the $E_{\alpha \alpha'}^{\mathbf{R} \mathbf{R'}}$ into two- and three-center integrals, following the guidelines of Ref. \onlinecite{slater_simplified_1954}.
As the explicit solution is straightforward, but very unhandy in print, it shall not be given here in full form. Instead, further details will be made accessible as supplementary material to this publication in mathematical notation in Ref. \onlinecite{supplementary_pdf} and, additionally, the explicit solution as MATLAB-compatible pseudocode in Ref. \onlinecite{supplementary_txt}, so that it can easily be used for own computations.

In order to give at least a brief insight into the physical meaning of the $E_{\alpha \alpha'}^{\mathbf{R} \mathbf{R'}}$, we will give the results of the expansion of Eq. (\ref{eq:EBOMmatrix}) to second order in $\mathbf{k}$. In this limit, the EBOM and the  $\mathbf{k}\cdot\mathbf{p}$-presentation become equivalent.
The following set of equations gives the EBOM matrix elements in terms of the parameters that were used in the 8-band-$\mathbf{k}\cdot\mathbf{p}$-Hamiltonian of Refs. \onlinecite{winkelnkemper_interrelation_2006} and \onlinecite{chuang_kp_1996}, where the $A_i$ are Luttinger-like parameters which are connected to the anisotropic effective valence band masses. For the sake of simplicity, the parameter $A_7$ has been set to zero in our approach. Its influence has turned out to be negligible.~\cite{dugdale_direct_2000}
The upper index in $E_{\alpha \alpha'}^{(k,l,m)}$  now denotes $\mathbf{R'}-\mathbf{R}$ in units of half the lattice constants $a$ or $c$, respectively, so that
\begin{displaymath}
 \mathbf{R'}-\mathbf{R} = \frac{k \, a}{2} \mathbf{e_x} + \frac{l \, a}{2} \mathbf{e_y} + \frac{m \, c}{2} \mathbf{e_z}.
\end{displaymath}
\begin{widetext}
\begin{eqnarray}
 \frac{\hbar}{2m_e^\parallel}-\frac{P_\parallel^2}{E_g} &=&\left(-4E_{ss}^{(\sqrt{3},1,2)}-2E_{ss}^{(0,2,2)}-4E_{ss}^{(0,0,4)}-E_{ss}^{(0,0,2)}\right)\cdot c^2 \nonumber,\\
 \frac{\hbar}{2m_e^\perp}-\frac{P_\perp^2}{E_g}&=&\left(-E_{ss}^{(0,2,0)}-E_{ss}^{(\sqrt{3},1,2)}-2E_{ss}^{(0,2,2)}-\frac{1}{2}E_{ss}^{(\sqrt{3},1,0)}-4E_{ss}^{(0,4,0)}-\frac{9}{2}E_{ss}^{(\sqrt{3},3,0)}-2E_{ss}^{(2\sqrt{3},2,0)}\right)\cdot a^2\nonumber,\\
 \frac{\hbar}{2m_e^\perp}-\frac{P_\perp^2}{E_g}&=&\left(-\frac{3}{2}E_{ss}^{(\sqrt{3},1,0)}-3E_{ss}^{(\sqrt{3},1,2)}-\frac{3}{2}E_{ss}^{(\sqrt{3},3,0)}-6E_{ss}^{(2\sqrt{3},2,0)}-3E_{ss}^{(2\sqrt{3},0,0)}\right)\cdot a^2\nonumber,\\
 iP_\perp&=&\left(2iE_{sx}^{(\sqrt{3},1,0)}+4iE_{sx}^{(\sqrt{3},1,2)}+2iE_{sx}^{(2\sqrt{3},0,0)}+4iE_{sx}^{(2\sqrt{3},2,0)}+2iE_{sx}^{(\sqrt{3},3,0)}\right)\cdot a\sqrt{3}\nonumber,\\
 iP_\perp&=&\left(2iE_{sy}^{(\sqrt{3},1,0)}+2iE_{sy}^{(0,2,0)}+4iE_{sy}^{(\sqrt{3},1,2)}+4iE_{sy}^{(0,2,2)}+4iE_{sy}^{(2\sqrt{3},2,0)}+6iE_{sy}^{(\sqrt{3},3,0)}+4iE_{sy}^{(0,4,0)}\right)\cdot a\nonumber,\\
 iP_\parallel&=&\left(2iE_{sz}^{(0,0,2)}+8iE_{sz}^{(\sqrt{3},1,2)}+4iE_{sz}^{(0,2,2)}+4iE_{sz}^{(0,0,4)}\right)\cdot c\nonumber,\\
 A_2+A_4+A_5+\frac{P_\parallel^2}{E_g}&=&\left(-\frac{3}{2}E_{xx}^{(\sqrt{3},1,0)}-3E_{xx}^{(\sqrt{3},1,2)}-\frac{3}{2}E_{xx}^{(\sqrt{3},3,0)}-6E_{xx}^{(2\sqrt{3},2,0)}-3E_{xx}^{(2\sqrt{3},0,0)}\right)\cdot a^2\nonumber,\\
 A_2+A_4-A_5&=&\left(-E_{xx}^{(0,2,0)}-E_{xx}^{(\sqrt{3},1,2)}-2E_{xx}^{(0,2,2)}-\frac{1}{2}E_{xx}^{(\sqrt{3},1,0)}-4E_{xx}^{(0,4,0)}-\frac{9}{2}E_{xx}^{(\sqrt{3},3,0)}-2E_{xx}^{(2\sqrt{3},2,0)}\right)\cdot a^2\nonumber,\\
 A_1+A_3&=&\left(-4E_{xx}^{(\sqrt{3},1,2)}-2E_{xx}^{(0,2,2)}-4E_{xx}^{(0,0,4)}-E_{xx}^{(0,0,2)}\right)\cdot c^2,\\
 2A_5+\frac{P_\parallel^2}{E_g}&=&\left(-E_{xy}^{(\sqrt{3},1,0)}-2E_{xy}^{(\sqrt{3},1,2)}-4E_{xy}^{(2\sqrt{3},2,0)}-3E_{xy}^{(\sqrt{3},3,0)}\right)\cdot a^2\sqrt{3}\nonumber,\\
 \sqrt{2}A_6+\frac{P_\parallel P_\perp}{E_g}&=&-4E_{xz}^{(\sqrt{3},1,2)}\cdot a\sqrt{3}c\nonumber,\\
 A_2+A_4-A_5&=&\left(-\frac{3}{2}E_{yy}^{(\sqrt{3},1,0)}-3E_{yy}^{(\sqrt{3},1,2)}-\frac{3}{2}E_{yy}^{(\sqrt{3},3,0)}-6E_{yy}^{(2\sqrt{3},2,0)}-3E_{yy}^{(2\sqrt{3},0,0)}\right)\cdot a^2\nonumber,\\
 A_2+A_4+A_5+\frac{P_\parallel^2}{E_g}&=&\left(-E_{yy}^{(0,2,0)}-E_{yy}^{(\sqrt{3},1,2)}-2E_{yy}^{(0,2,2)}-\frac{1}{2}E_{yy}^{(\sqrt{3},1,0)}-4E_{yy}^{(0,4,0)}-\frac{9}{2}E_{yy}^{(\sqrt{3},3,0)}-2E_{yy}^{(2\sqrt{3},2,0)}\right)\cdot a^2\nonumber,\\
 A_1+A_3&=&\left(-4E_{yy}^{(\sqrt{3},1,2)}-2E_{yy}^{(0,2,2)}-4E_{yy}^{(0,0,4)}-E_{yy}^{(0,0,2)}\right)\cdot c^2\nonumber,\\
 \sqrt{2}A_6+\frac{P_\parallel P_\perp}{E_g}&=&\left(-4E_{yz}^{(\sqrt{3},1,2)}-4E_{yz}^{(0,2,2)}\right)\cdot ac\nonumber,\\
 A_2&=&\left(-\frac{3}{2}E_{zz}^{(\sqrt{3},1,0)}-3E_{zz}^{(\sqrt{3},1,2)}-\frac{3}{2}E_{zz}^{(\sqrt{3},3,0)}-6E_{zz}^{(2\sqrt{3},2,0)}-3E_{zz}^{(2\sqrt{3},0,0)}\right)\cdot a^2\nonumber,\\
 A_2&=&\left(-E_{zz}^{(0,2,0)}-E_{zz}^{(\sqrt{3},1,2)}-2E_{zz}^{(0,2,2)}-\frac{1}{2}E_{zz}^{(\sqrt{3},1,0)}-4E_{zz}^{(0,4,0)}-\frac{9}{2}E_{zz}^{(\sqrt{3},3,0)}-2E_{zz}^{(2\sqrt{3},2,0)}\right)\cdot a^2\nonumber,\\
 A_1+\frac{P_\perp^2}{E_g}&=&\left(-4E_{zz}^{(\sqrt{3},1,2)}-2E_{zz}^{(0,2,2)}-4E_{zz}^{(0,0,4)}-E_{zz}^{(0,0,2)}\right)\cdot c^2 \nonumber.
\end{eqnarray}
\end{widetext}
%
 

\begin{thebibliography}{45}
\expandafter\ifx\csname natexlab\endcsname\relax\def\natexlab#1{#1}\fi
\expandafter\ifx\csname bibnamefont\endcsname\relax
  \def\bibnamefont#1{#1}\fi
\expandafter\ifx\csname bibfnamefont\endcsname\relax
  \def\bibfnamefont#1{#1}\fi
\expandafter\ifx\csname citenamefont\endcsname\relax
  \def\citenamefont#1{#1}\fi
\expandafter\ifx\csname url\endcsname\relax
  \def\url#1{\texttt{#1}}\fi
\expandafter\ifx\csname urlprefix\endcsname\relax\def\urlprefix{URL }\fi
\providecommand{\bibinfo}[2]{#2}
\providecommand{\eprint}[2][]{\url{#2}}

\bibitem[{\citenamefont{Guzelian et~al.}(1996)\citenamefont{Guzelian, Banin,
  Kadavanich, Peng, and Alivisatos}}]{guzelian_colloidal_1996}
\bibinfo{author}{\bibfnamefont{A.~A.} \bibnamefont{Guzelian}},
  \bibinfo{author}{\bibfnamefont{U.}~\bibnamefont{Banin}},
  \bibinfo{author}{\bibfnamefont{A.~V.} \bibnamefont{Kadavanich}},
  \bibinfo{author}{\bibfnamefont{X.}~\bibnamefont{Peng}}, \bibnamefont{and}
  \bibinfo{author}{\bibfnamefont{A.~P.} \bibnamefont{Alivisatos}},
  \bibinfo{journal}{Appl. Phys. Lett.} \textbf{\bibinfo{volume}{69}},
  \bibinfo{pages}{1432} (\bibinfo{year}{1996}).

\bibitem[{\citenamefont{Michler}(2009)}]{michler_single_2009}
\bibinfo{author}{\bibfnamefont{P.}~\bibnamefont{Michler}},
  \emph{\bibinfo{title}{Single Semiconductor Quantum Dots}}
  (\bibinfo{publisher}{Springer}, \bibinfo{year}{2009}), \bibinfo{edition}{1st}
  ed.

\bibitem[{\citenamefont{Bruchez et~al.}(1998)\citenamefont{Bruchez, Moronne,
  Gin, Weiss, and Alivisatos}}]{bruchez_semiconductor_1998}
\bibinfo{author}{\bibfnamefont{M.}~\bibnamefont{Bruchez}},
  \bibinfo{author}{\bibfnamefont{M.}~\bibnamefont{Moronne}},
  \bibinfo{author}{\bibfnamefont{P.}~\bibnamefont{Gin}},
  \bibinfo{author}{\bibfnamefont{S.}~\bibnamefont{Weiss}}, \bibnamefont{and}
  \bibinfo{author}{\bibfnamefont{A.~P.} \bibnamefont{Alivisatos}},
  \bibinfo{journal}{Science} \textbf{\bibinfo{volume}{281}},
  \bibinfo{pages}{2013} (\bibinfo{year}{1998}).

\bibitem[{\citenamefont{Michalet et~al.}(2005)\citenamefont{Michalet, Pinaud,
  Bentolila, Tsay, Doose, Li, Sundaresan, Wu, Gambhir, and
  Weiss}}]{michalet_quantum_2005}
\bibinfo{author}{\bibfnamefont{X.}~\bibnamefont{Michalet}},
  \bibinfo{author}{\bibfnamefont{F.~F.} \bibnamefont{Pinaud}},
  \bibinfo{author}{\bibfnamefont{L.~A.} \bibnamefont{Bentolila}},
  \bibinfo{author}{\bibfnamefont{J.~M.} \bibnamefont{Tsay}},
  \bibinfo{author}{\bibfnamefont{S.}~\bibnamefont{Doose}},
  \bibinfo{author}{\bibfnamefont{J.~J.} \bibnamefont{Li}},
  \bibinfo{author}{\bibfnamefont{G.}~\bibnamefont{Sundaresan}},
  \bibinfo{author}{\bibfnamefont{A.~M.} \bibnamefont{Wu}},
  \bibinfo{author}{\bibfnamefont{S.~S.} \bibnamefont{Gambhir}},
  \bibnamefont{and} \bibinfo{author}{\bibfnamefont{S.}~\bibnamefont{Weiss}},
  \bibinfo{journal}{Science} \textbf{\bibinfo{volume}{307}},
  \bibinfo{pages}{538} (\bibinfo{year}{2005}).

\bibitem[{\citenamefont{Lazar et~al.}(2004)\citenamefont{Lazar, H\'ebert, and
  Zandbergen}}]{lazar_investigation_2004}
\bibinfo{author}{\bibfnamefont{S.}~\bibnamefont{Lazar}},
  \bibinfo{author}{\bibfnamefont{C.}~\bibnamefont{H\'ebert}}, \bibnamefont{and}
  \bibinfo{author}{\bibfnamefont{H.~W.} \bibnamefont{Zandbergen}},
  \bibinfo{journal}{Ultramicroscopy} \textbf{\bibinfo{volume}{98}},
  \bibinfo{pages}{249} (\bibinfo{year}{2004}).

\bibitem[{\citenamefont{Grundmann et~al.}(1995)\citenamefont{Grundmann, Stier,
  and Bimberg}}]{grundmann_inas/gaas_1995}
\bibinfo{author}{\bibfnamefont{M.}~\bibnamefont{Grundmann}},
  \bibinfo{author}{\bibfnamefont{O.}~\bibnamefont{Stier}}, \bibnamefont{and}
  \bibinfo{author}{\bibfnamefont{D.}~\bibnamefont{Bimberg}},
  \bibinfo{journal}{Phys. Rev. B} \textbf{\bibinfo{volume}{52}},
  \bibinfo{pages}{11969} (\bibinfo{year}{1995}).

\bibitem[{\citenamefont{Wojs et~al.}(1996)\citenamefont{Wojs, Hawrylak, Fafard,
  and Jacak}}]{wojs_electronic_1996}
\bibinfo{author}{\bibfnamefont{A.}~\bibnamefont{Wojs}},
  \bibinfo{author}{\bibfnamefont{P.}~\bibnamefont{Hawrylak}},
  \bibinfo{author}{\bibfnamefont{S.}~\bibnamefont{Fafard}}, \bibnamefont{and}
  \bibinfo{author}{\bibfnamefont{L.}~\bibnamefont{Jacak}},
  \bibinfo{journal}{Phys. Rev. B} \textbf{\bibinfo{volume}{54}},
  \bibinfo{pages}{5604} (\bibinfo{year}{1996}).

\bibitem[{\citenamefont{Shi and Gan}(2003)}]{shi_effects_2003}
\bibinfo{author}{\bibfnamefont{J.}~\bibnamefont{Shi}} \bibnamefont{and}
  \bibinfo{author}{\bibfnamefont{Z.}~\bibnamefont{Gan}}, \bibinfo{journal}{J.
  Appl. Phys.} \textbf{\bibinfo{volume}{94}}, \bibinfo{pages}{407}
  (\bibinfo{year}{2003}).

\bibitem[{\citenamefont{Fonoberov and
  Balandin}(2003)}]{fonoberov_excitonic_2003}
\bibinfo{author}{\bibfnamefont{V.~A.} \bibnamefont{Fonoberov}}
  \bibnamefont{and} \bibinfo{author}{\bibfnamefont{A.~A.}
  \bibnamefont{Balandin}}, \bibinfo{journal}{J. Appl. Phys.}
  \textbf{\bibinfo{volume}{94}}, \bibinfo{pages}{7178} (\bibinfo{year}{2003}).

\bibitem[{\citenamefont{Pryor}(1998)}]{pryor_eight-band_1998}
\bibinfo{author}{\bibfnamefont{C.}~\bibnamefont{Pryor}},
  \bibinfo{journal}{Phys. Rev. B} \textbf{\bibinfo{volume}{57}},
  \bibinfo{pages}{7190} (\bibinfo{year}{1998}).

\bibitem[{\citenamefont{Stier et~al.}(1999)\citenamefont{Stier, Grundmann, and
  Bimberg}}]{stier_electronic_1999}
\bibinfo{author}{\bibfnamefont{O.}~\bibnamefont{Stier}},
  \bibinfo{author}{\bibfnamefont{M.}~\bibnamefont{Grundmann}},
  \bibnamefont{and} \bibinfo{author}{\bibfnamefont{D.}~\bibnamefont{Bimberg}},
  \bibinfo{journal}{Phys. Rev. B} \textbf{\bibinfo{volume}{59}},
  \bibinfo{pages}{5688} (\bibinfo{year}{1999}).

\bibitem[{\citenamefont{Andreev and {O’Reilly}}(2000)}]{andreev_theory_2000}
\bibinfo{author}{\bibfnamefont{A.~D.} \bibnamefont{Andreev}} \bibnamefont{and}
  \bibinfo{author}{\bibfnamefont{E.~P.} \bibnamefont{{O’Reilly}}},
  \bibinfo{journal}{Phys. Rev. B} \textbf{\bibinfo{volume}{62}},
  \bibinfo{pages}{15851} (\bibinfo{year}{2000}).

\bibitem[{\citenamefont{Wang and Zunger}(1996)}]{wang_pseudopotential_1996}
\bibinfo{author}{\bibfnamefont{L.~W.}~\bibnamefont{Wang}} \bibnamefont{and}
  \bibinfo{author}{\bibfnamefont{A.}~\bibnamefont{Zunger}},
  \bibinfo{journal}{Phys. Rev. B} \textbf{\bibinfo{volume}{53}},
  \bibinfo{pages}{9579} (\bibinfo{year}{1996}).

\bibitem[{\citenamefont{Wang and Zunger}(1999)}]{wang_linear_1999}
\bibinfo{author}{\bibfnamefont{L.~W.}~\bibnamefont{Wang}} \bibnamefont{and}
  \bibinfo{author}{\bibfnamefont{A.}~\bibnamefont{Zunger}},
  \bibinfo{journal}{Phys. Rev. B} \textbf{\bibinfo{volume}{59}},
  \bibinfo{pages}{15806} (\bibinfo{year}{1999}).

\bibitem[{\citenamefont{Wang et~al.}(2000)\citenamefont{Wang, Williamson,
  Zunger, Jiang, and Singh}}]{wang_comparison_2000}
\bibinfo{author}{\bibfnamefont{L.~W.} \bibnamefont{Wang}},
  \bibinfo{author}{\bibfnamefont{A.~J.} \bibnamefont{Williamson}},
  \bibinfo{author}{\bibfnamefont{A.}~\bibnamefont{Zunger}},
  \bibinfo{author}{\bibfnamefont{H.}~\bibnamefont{Jiang}}, \bibnamefont{and}
  \bibinfo{author}{\bibfnamefont{J.}~\bibnamefont{Singh}},
  \bibinfo{journal}{Appl. Phys. Lett.} \textbf{\bibinfo{volume}{76}},
  \bibinfo{pages}{339} (\bibinfo{year}{2000}).

\bibitem[{\citenamefont{Bester and Zunger}(2005)}]{bester_cylindrically_2005}
\bibinfo{author}{\bibfnamefont{G.}~\bibnamefont{Bester}} \bibnamefont{and}
  \bibinfo{author}{\bibfnamefont{A.}~\bibnamefont{Zunger}},
  \bibinfo{journal}{Phys. Rev. B} \textbf{\bibinfo{volume}{71}},
  \bibinfo{pages}{045318} (\bibinfo{year}{2005}).

\bibitem[{\citenamefont{Santoprete et~al.}(2003)\citenamefont{Santoprete,
  Koiller, Capaz, Kratzer, Liu, and Scheffler}}]{santoprete_tight-binding_2003}
\bibinfo{author}{\bibfnamefont{R.}~\bibnamefont{Santoprete}},
  \bibinfo{author}{\bibfnamefont{B.}~\bibnamefont{Koiller}},
  \bibinfo{author}{\bibfnamefont{R.~B.} \bibnamefont{Capaz}},
  \bibinfo{author}{\bibfnamefont{P.}~\bibnamefont{Kratzer}},
  \bibinfo{author}{\bibfnamefont{Q.~K.~K.} \bibnamefont{Liu}},
  \bibnamefont{and}
  \bibinfo{author}{\bibfnamefont{M.}~\bibnamefont{Scheffler}},
  \bibinfo{journal}{Phys. Rev. B} \textbf{\bibinfo{volume}{68}},
  \bibinfo{pages}{235311} (\bibinfo{year}{2003}).

\bibitem[{\citenamefont{Schulz and Czycholl}(2005)}]{schulz_tight-binding_2005}
\bibinfo{author}{\bibfnamefont{S.}~\bibnamefont{Schulz}} \bibnamefont{and}
  \bibinfo{author}{\bibfnamefont{G.}~\bibnamefont{Czycholl}},
  \bibinfo{journal}{Phys. Rev. B} \textbf{\bibinfo{volume}{72}},
  \bibinfo{pages}{165317} (\bibinfo{year}{2005}).

\bibitem[{\citenamefont{Schulz et~al.}(2006)\citenamefont{Schulz, Schumacher,
  and Czycholl}}]{schulz_tight-binding_2006}
\bibinfo{author}{\bibfnamefont{S.}~\bibnamefont{Schulz}},
  \bibinfo{author}{\bibfnamefont{S.}~\bibnamefont{Schumacher}},
  \bibnamefont{and} \bibinfo{author}{\bibfnamefont{G.}~\bibnamefont{Czycholl}},
  \bibinfo{journal}{Phys. Rev. B} \textbf{\bibinfo{volume}{73}},
  \bibinfo{pages}{245327} (\bibinfo{year}{2006}).

\bibitem[{\citenamefont{Schulz and Czycholl}(2006)}]{schulz_electronic_2006}
\bibinfo{author}{\bibfnamefont{S.}~\bibnamefont{Schulz}} \bibnamefont{and}
  \bibinfo{author}{\bibfnamefont{G.}~\bibnamefont{Czycholl}},
  \bibinfo{journal}{phys. stat. sol. (c)} \textbf{\bibinfo{volume}{3}},
  \bibinfo{pages}{1675} (\bibinfo{year}{2006}).


\bibitem[{\citenamefont{Schulz et~al.}(2008)\citenamefont{Schulz, Schumacher,
  and Czycholl}}]{schulz_spin-orbit_2008}
\bibinfo{author}{\bibfnamefont{S.}~\bibnamefont{Schulz}},
  \bibinfo{author}{\bibfnamefont{S.}~\bibnamefont{Schumacher}},
  \bibnamefont{and} \bibinfo{author}{\bibfnamefont{G.}~\bibnamefont{Czycholl}},
  \bibinfo{journal}{The European Physical Journal B}
  \textbf{\bibinfo{volume}{64}}, \bibinfo{pages}{51} (\bibinfo{year}{2008}).

\bibitem[{\citenamefont{Korkusinski et~al.}(2008)\citenamefont{Korkusinski,
  Hawrylak, Zielinski, Sheng, and Klimeck}}]{korkusinski_building_2008}
\bibinfo{author}{\bibfnamefont{M.}~\bibnamefont{Korkusinski}},
  \bibinfo{author}{\bibfnamefont{P.}~\bibnamefont{Hawrylak}},
  \bibinfo{author}{\bibfnamefont{M.}~\bibnamefont{Zielinski}},
  \bibinfo{author}{\bibfnamefont{W.}~\bibnamefont{Sheng}}, \bibnamefont{and}
  \bibinfo{author}{\bibfnamefont{G.}~\bibnamefont{Klimeck}},
  \bibinfo{journal}{Microelectronics Journal} \textbf{\bibinfo{volume}{39}},
  \bibinfo{pages}{318} (\bibinfo{year}{2008}).

\bibitem[{\citenamefont{Rinke et~al.}(2006)\citenamefont{Rinke, Scheffler,
  Qteish, Winkelnkemper, Bimberg, and Neugebauer}}]{rinke_band_2006}
\bibinfo{author}{\bibfnamefont{P.}~\bibnamefont{Rinke}},
  \bibinfo{author}{\bibfnamefont{M.}~\bibnamefont{Scheffler}},
  \bibinfo{author}{\bibfnamefont{A.}~\bibnamefont{Qteish}},
  \bibinfo{author}{\bibfnamefont{M.}~\bibnamefont{Winkelnkemper}},
  \bibinfo{author}{\bibfnamefont{D.}~\bibnamefont{Bimberg}}, \bibnamefont{and}
  \bibinfo{author}{\bibfnamefont{J.}~\bibnamefont{Neugebauer}},
  \bibinfo{journal}{Appl. Phys. Lett.} \textbf{\bibinfo{volume}{89}},
  \bibinfo{pages}{161919} (\bibinfo{year}{2006}).

\bibitem[{\citenamefont{Rinke et~al.}(2008)\citenamefont{Rinke, Winkelnkemper,
  Qteish, Bimberg, Neugebauer, and Scheffler}}]{rinke_consistent_2008}
\bibinfo{author}{\bibfnamefont{P.}~\bibnamefont{Rinke}},
  \bibinfo{author}{\bibfnamefont{M.}~\bibnamefont{Winkelnkemper}},
  \bibinfo{author}{\bibfnamefont{A.}~\bibnamefont{Qteish}},
  \bibinfo{author}{\bibfnamefont{D.}~\bibnamefont{Bimberg}},
  \bibinfo{author}{\bibfnamefont{J.}~\bibnamefont{Neugebauer}},
  \bibnamefont{and}
  \bibinfo{author}{\bibfnamefont{M.}~\bibnamefont{Scheffler}},
  \bibinfo{journal}{Phys. Rev. B} \textbf{\bibinfo{volume}{77}},
  \bibinfo{pages}{075202} (\bibinfo{year}{2008}).

\bibitem[{\citenamefont{Slater and Koster}(1954)}]{slater_simplified_1954}
\bibinfo{author}{\bibfnamefont{J.~C.} \bibnamefont{Slater}} \bibnamefont{and}
  \bibinfo{author}{\bibfnamefont{G.~F.} \bibnamefont{Koster}},
  \bibinfo{journal}{Phys. Rev.} \textbf{\bibinfo{volume}{94}},
  \bibinfo{pages}{1498} (\bibinfo{year}{1954}).

\bibitem[{\citenamefont{Chang}(1988)}]{chang_bond-orbital_1988}
\bibinfo{author}{\bibfnamefont{Y.~C.}~\bibnamefont{Chang}},
  \bibinfo{journal}{Phys. Rev. B} \textbf{\bibinfo{volume}{37}},
  \bibinfo{pages}{8215} (\bibinfo{year}{1988}).

\bibitem[{\citenamefont{Loehr}(1994)}]{loehr_improved_1994}
\bibinfo{author}{\bibfnamefont{J.~P.} \bibnamefont{Loehr}},
  \bibinfo{journal}{Phys. Rev. B} \textbf{\bibinfo{volume}{50}},
  \bibinfo{pages}{5429} (\bibinfo{year}{1994}).

\bibitem[{\citenamefont{Fritsch et~al.}(2004)\citenamefont{Fritsch, Schmidt,
  and Grundmann}}]{fritsch_band_2004}
\bibinfo{author}{\bibfnamefont{D.}~\bibnamefont{Fritsch}},
  \bibinfo{author}{\bibfnamefont{H.}~\bibnamefont{Schmidt}}, \bibnamefont{and}
  \bibinfo{author}{\bibfnamefont{M.}~\bibnamefont{Grundmann}},
  \bibinfo{journal}{Phys. Rev. B} \textbf{\bibinfo{volume}{69}},
  \bibinfo{pages}{165204} (\bibinfo{year}{2004}).

\bibitem[{\citenamefont{Chen}(2004)}]{chen_bond_2004}
\bibinfo{author}{\bibfnamefont{C.}~\bibnamefont{Chen}},
  \bibinfo{journal}{Physics Letters A} \textbf{\bibinfo{volume}{329}},
  \bibinfo{pages}{136} (\bibinfo{year}{2004}).

\bibitem[{\citenamefont{Cartoix\'a et~al.}(2003)\citenamefont{Cartoix\'a, Ting,
  and {McGill}}}]{cartoixa_description_2003}
\bibinfo{author}{\bibfnamefont{X.}~\bibnamefont{Cartoix\'a}},
  \bibinfo{author}{\bibfnamefont{D.~Z.~-Y.} \bibnamefont{Ting}}, \bibnamefont{and}
  \bibinfo{author}{\bibfnamefont{T.~C.} \bibnamefont{{McGill}}},
  \bibinfo{journal}{Phys. Rev. B} \textbf{\bibinfo{volume}{68}},
  \bibinfo{pages}{235319} (\bibinfo{year}{2003}).

\bibitem[{\citenamefont{Chadi}(1977)}]{chadi_spin-orbit_1977}
\bibinfo{author}{\bibfnamefont{D.~J.} \bibnamefont{Chadi}},
  \bibinfo{journal}{Phys. Rev. B} \textbf{\bibinfo{volume}{16}},
  \bibinfo{pages}{790} (\bibinfo{year}{1977}).

\bibitem[{\citenamefont{Winkelnkemper et~al.}(2006)\citenamefont{Winkelnkemper,
  Schliwa, and Bimberg}}]{winkelnkemper_interrelation_2006}
\bibinfo{author}{\bibfnamefont{M.}~\bibnamefont{Winkelnkemper}},
  \bibinfo{author}{\bibfnamefont{A.}~\bibnamefont{Schliwa}}, \bibnamefont{and}
  \bibinfo{author}{\bibfnamefont{D.}~\bibnamefont{Bimberg}},
  \bibinfo{journal}{Phys. Rev. B} \textbf{\bibinfo{volume}{74}},
  \bibinfo{pages}{155322} (\bibinfo{year}{2006}).

\bibitem[{\citenamefont{Chuang and Chang}(1996)}]{chuang_kp_1996}
\bibinfo{author}{\bibfnamefont{S.~L.} \bibnamefont{Chuang}} \bibnamefont{and}
  \bibinfo{author}{\bibfnamefont{C.~S.} \bibnamefont{Chang}},
  \bibinfo{journal}{Phys. Rev. B} \textbf{\bibinfo{volume}{54}},
  \bibinfo{pages}{2491} (\bibinfo{year}{1996}).

\bibitem[{\citenamefont{Yeo et~al.}(1998)\citenamefont{Yeo, Chong, and
  Li}}]{yeo_electronic_1998}
\bibinfo{author}{\bibfnamefont{Y.~C.} \bibnamefont{Yeo}},
  \bibinfo{author}{\bibfnamefont{T.~C.} \bibnamefont{Chong}}, \bibnamefont{and}
  \bibinfo{author}{\bibfnamefont{M.~F.} \bibnamefont{Li}}, \bibinfo{journal}{J.
  Appl. Phys.} \textbf{\bibinfo{volume}{83}}, \bibinfo{pages}{1429}
  (\bibinfo{year}{1998}).

\bibitem[{\citenamefont{Rinke et~al.}(2005)\citenamefont{Rinke, Qteish,
  Neugebauer, Freysoldt, and Scheffler}}]{rinke_combining_2005}
\bibinfo{author}{\bibfnamefont{P.}~\bibnamefont{Rinke}},
  \bibinfo{author}{\bibfnamefont{A.}~\bibnamefont{Qteish}},
  \bibinfo{author}{\bibfnamefont{J.}~\bibnamefont{Neugebauer}},
  \bibinfo{author}{\bibfnamefont{C.}~\bibnamefont{Freysoldt}},
  \bibnamefont{and}
  \bibinfo{author}{\bibfnamefont{M.}~\bibnamefont{Scheffler}},
  \bibinfo{journal}{New Journal of Physics} \textbf{\bibinfo{volume}{7}},
  \bibinfo{pages}{126} (\bibinfo{year}{2005}), ISSN \bibinfo{issn}{1367-2630}.

\bibitem[{\citenamefont{Rinke}(2009)}]{rinke_personal_2009}
\bibinfo{author}{\bibfnamefont{P.}~\bibnamefont{Rinke}},
  \bibinfo{journal}{obtained by private communication}  (\bibinfo{year}{2009}).

\bibitem[{\citenamefont{Vurgaftman and Meyer}(2003)}]{vurgaftman_band_2003}
\bibinfo{author}{\bibfnamefont{I.}~\bibnamefont{Vurgaftman}} \bibnamefont{and}
  \bibinfo{author}{\bibfnamefont{J.~R.} \bibnamefont{Meyer}},
  \bibinfo{journal}{J. Appl. Phys.} \textbf{\bibinfo{volume}{94}},
  \bibinfo{pages}{3675} (\bibinfo{year}{2003}).

\bibitem[{\citenamefont{Winkelnkemper et~al.}(2007)\citenamefont{Winkelnkemper,
  Seguin, Rodt, Schliwa, Reissmann, Strittmatter, Hoffmann, and
  Bimberg}}]{winkelnkemper_polarized_2007}
\bibinfo{author}{\bibfnamefont{M.}~\bibnamefont{Winkelnkemper}},
  \bibinfo{author}{\bibfnamefont{R.}~\bibnamefont{Seguin}},
  \bibinfo{author}{\bibfnamefont{S.}~\bibnamefont{Rodt}},
  \bibinfo{author}{\bibfnamefont{A.}~\bibnamefont{Schliwa}},
  \bibinfo{author}{\bibfnamefont{L.}~\bibnamefont{Reissmann}},
  \bibinfo{author}{\bibfnamefont{A.}~\bibnamefont{Strittmatter}},
  \bibinfo{author}{\bibfnamefont{A.}~\bibnamefont{Hoffmann}}, \bibnamefont{and}
  \bibinfo{author}{\bibfnamefont{D.}~\bibnamefont{Bimberg}},
  \bibinfo{journal}{J. Appl. Phys.} \textbf{\bibinfo{volume}{101}},
  \bibinfo{pages}{113708} (\bibinfo{year}{2007}).

\bibitem[{\citenamefont{Wang and Zunger}(1994)}]{wang_solving_1994}
\bibinfo{author}{\bibfnamefont{L.}~\bibnamefont{Wang}} \bibnamefont{and}
  \bibinfo{author}{\bibfnamefont{A.}~\bibnamefont{Zunger}},
  \bibinfo{journal}{The Journal of Chemical Physics}
  \textbf{\bibinfo{volume}{100}}, \bibinfo{pages}{2394} (\bibinfo{year}{1994}).

\bibitem[{\citenamefont{Wei and Zunger}(1996)}]{wei_valence_1996}
\bibinfo{author}{\bibfnamefont{S.}~\bibnamefont{Wei}} \bibnamefont{and}
  \bibinfo{author}{\bibfnamefont{A.}~\bibnamefont{Zunger}},
  \bibinfo{journal}{Appl. Phys. Lett.} \textbf{\bibinfo{volume}{69}},
  \bibinfo{pages}{2719} (\bibinfo{year}{1996}).

\bibitem[{\citenamefont{Marquardt et~al.}(2008)\citenamefont{Marquardt, Mourad,
  Schulz, Hickel, Czycholl, and Neugebauer}}]{marquardt_comparison_2008}
\bibinfo{author}{\bibfnamefont{O.}~\bibnamefont{Marquardt}},
  \bibinfo{author}{\bibfnamefont{D.}~\bibnamefont{Mourad}},
  \bibinfo{author}{\bibfnamefont{S.}~\bibnamefont{Schulz}},
  \bibinfo{author}{\bibfnamefont{T.}~\bibnamefont{Hickel}},
  \bibinfo{author}{\bibfnamefont{G.}~\bibnamefont{Czycholl}}, \bibnamefont{and}
  \bibinfo{author}{\bibfnamefont{J.}~\bibnamefont{Neugebauer}},
  \bibinfo{journal}{Phys. Rev. B} \textbf{\bibinfo{volume}{78}},
  \bibinfo{pages}{235302} (\bibinfo{year}{2008}).

\bibitem[{\citenamefont{Schulz et~al.}(2009)\citenamefont{Schulz, Mourad, and
  Czycholl}}]{schulz_multiband_2009}
\bibinfo{author}{\bibfnamefont{S.}~\bibnamefont{Schulz}},
  \bibinfo{author}{\bibfnamefont{D.}~\bibnamefont{Mourad}}, \bibnamefont{and}
  \bibinfo{author}{\bibfnamefont{G.}~\bibnamefont{Czycholl}},
  \bibinfo{journal}{Phys. Rev. B} \textbf{\bibinfo{volume}{80}},
  \bibinfo{pages}{165405} (\bibinfo{year}{2009}).


\bibitem[{sup(2010{\natexlab{a}})}]{supplementary_pdf}
\bibinfo{journal}{See supplementary document supplementary\_material.pdf for further details of the
  parametrization}.

\bibitem[{sup(2010{\natexlab{b}})}]{supplementary_txt}
\bibinfo{journal}{See supplementary file ebom\_integrals\_snn.txt for the explicit solution of the
  resulting system of equations}.


\bibitem[{\citenamefont{Dugdale et~al.}(2000)\citenamefont{Dugdale, Brand, and
  Abram}}]{dugdale_direct_2000}
\bibinfo{author}{\bibfnamefont{D.~J.} \bibnamefont{Dugdale}},
  \bibinfo{author}{\bibfnamefont{S.}~\bibnamefont{Brand}}, \bibnamefont{and}
  \bibinfo{author}{\bibfnamefont{R.~A.} \bibnamefont{Abram}},
  \bibinfo{journal}{Phys. Rev. B} \textbf{\bibinfo{volume}{61}},
  \bibinfo{pages}{12933} (\bibinfo{year}{2000}).

\bibitem[{\citenamefont{Froyen and Harrison}(1979)}]{froyen_elementary_1979}
\bibinfo{author}{\bibfnamefont{S.}~\bibnamefont{Froyen}} \bibnamefont{and}
  \bibinfo{author}{\bibfnamefont{W.~A.}~\bibnamefont{Harrison}},
  \bibinfo{journal}{Phys. Rev. B} \textbf{\bibinfo{volume}{20}},
  \bibinfo{pages}{2420} (\bibinfo{year}{1979}).

\end{thebibliography}

\end{document}